\documentclass{PoS}
\usepackage{wrapfig}
\usepackage{subfigure}
\usepackage{graphicx}
\title{Fragmentation Functions measurement at COMPASS}

\ShortTitle{Fragmentation Functions measurement at COMPASS}

\author{\speaker{Nour MAKKE}
       ( Universita e INFN ( IT))\\
       E-mail: \email{nour.makke@cern.ch}}

\abstract{Fragmentation functions represent a key ingredient to address the proton spin structure in semi-inclusive deep-inelastic scattering and proton-proton collisions. They can not be determined from perturbative Quantum Chromodynamics and have to be extracted from experimental data in different processes. The COMPASS experiment at CERN provides a large data sample and covers a wide kinematic range for precise measurement of hadron multiplicities, directly connected to fragmentation functions. Recent,  full-differential results on pion and kaon multiplicities are presented and discussed.}

\FullConference{XXI International Workshop on Deep-Inelastic Scattering and Related Subjects\\
		 22-26 April, 2013\\
		 Marseilles, France}

\begin{document}

\section{Introduction}

Semi-inclusive lepton-nucleon deep-inelastic scattering (SIDIS) is one of the most powerful tools to study simultaneously the internal structure of the nucleon and the quark hadronisation mechanism. It is described, according to the factorisation theorem, by a hard lepton-parton scattering cross section, calculable in perturbative chromodynamics (pQCD), convoluted with universal non-perturbative Parton Distribution Functions (PDFs) and Fragmentation Functions (FFs). The fragmentation functions encode the details of the hadronisation mechanism which describes the transition of a parton into a hadron in the final state. In other terms, the FFs reflect the number density of final-state hadrons produced in the hadronisation of partons. While PDFs are nowadays precisely known thanks to the enormous scientific effort in the last decade, the hadronisation mechanisms remains at a very preliminary stage of knowledge with a growing interest in more accurate and precise measurements of fragmentation functions. Two aspects of this mechanism are important to be studied: single-hadron (FFs) and hadron pair (DFFs) fragmentation functions. While the FFs represent a key element in the flavor separation of polarised PDFs and play a particular role in the $\Delta S$ puzzle (described in sec.~\ref{DSpuzzle}), the DFFs are crucial to access the transversity, a poorly known cornerstone in the nucleon spin structure \cite{Adolph:2012nw}. Only FFs are discussed here (for DFFs, new results were presented in~\cite{MakkeSPIN12}). The current fragmentation region, which represents the hadronisation of the struck quark, is investigated.

Currently, the flavor dependence of FFs is a crucial tool for a complete picture of the nucleon longitudinal spin structure. Our current knowledge of single hadron fragmentation functions is based on existing global QCD analyses mainly driven by high precision measurements from single inclusive electron-positron annihilation into charged hadrons, for which the cross section has no dependence on parton distribution functions, making it a clean process to study the hadronisation mechanism. However these data do not disentangle between quark and anti-quark fragmentation. In addition, most of the existing measurements at LEP and SLAC, are performed at the hard scale of the $Z$ boson mass. At this scale, for e$^+$e$^-$ data, electroweak coupling constants become equal and only the singlet combination of FFs can be extracted. Recently, a complementary precision measurement of inclusive e$^+$e$^-$ annihilation at the hard scale $Q^{2} = 10.52$ (GeV)$^2$ was reported by the BELLE collaboration, which is expected to improve the precision of QCD global analyses of fragmentation functions. In a recent QCD global fit of world data, results on transverse-momentum spectra of single-hadron  inclusively produced in proton-proton collisions at central and forward rapidities at RHIC have been included allowing for a better constraint of the gluon fragmentation function.  Complementary information on the parton fragmentation in a complementary energy regime and for different flavors is provided by semi-inclusive measurements in lepton-nucleon deep-inelastic scattering. In this case, the measurement is performed at lower hard scales than those of the collider. In addition it covers a wide hard scale range and provides a unique way to investigate the flavor structure of fragmentation functions. Note that the fragmentation of a quark of a specific flavor into a final-state hadron is called favoured if the quark is one of its valence quarks and in the other case, the fragmentation function is called unfavoured. Nowadays, pion fragmentation functions are known with a limited precision while kaon fragmentation functions are poorly known. 

The semi-inclusive $\gamma(q) N(P) \rightarrow \textit{h}(p_{h})X$ reaction, where $q$, $P$ and $p_{h}$ denote the four-momenta of the virtual photon $\gamma$, the nucleon $N$ and the observed hadron $h$, while $X$ denotes unobserved hadrons, is described by the virtual photon transfer four-momentum squared $Q^{2}= -q^{2}$ and the Bjorken variable $x = -q^{2}/(2P\cdot q)$. One can further define two important variables: the lepton energy fraction carried by the virtual photon $y$, the invariant mass of the final hadronic system $W = \sqrt{(P + q)^{2}}$ and the virtual photon energy fraction carried by final-state hadron $z = (P\cdot p_{h})/(P\cdot q)$.  The most relevant SIDIS observable to study the hadronisation process is the hadron multiplicity shown at LO in Eq. \ref{MulDef} for the single-hadron case. It depends simultaneously upon the PDFs ($q(x,Q^{2}$)) and the FFs ($D_{q}^{h}(z,Q^{2})$). 

\begin{equation}
M^{h}(x,Q^{2},z) = \frac{d^{3}\sigma^{h}(x,Q^{2},z)/dxdQ^{2}dz}{d^{2}\sigma^{DIS}(x,Q^{2})/dxdQ^{2}} = \frac{\sum_{q}e_{q}^{2}q(x,Q^{2}) \cdot D_{q}^{h}(z,Q^{2})}{\sum_{q} e_{q}^{2}q(x,Q^{2})}
\label{MulDef}
\end{equation}

In semi-inclusive DIS, the FFs are accessed by measuring hadron multiplicities which are defined by the averaged number of hadrons produced per DIS event. The experimental multiplicities must then be normalized by the acceptance correction factor which takes into account the limited angular and geometrical acceptance covered by the experimental apparatus and kinematic smearing and finally corrected for radiative effects.

\section{The $\Delta S$ puzzle}
\label{DSpuzzle}
The first moment of the strange quark polarisation in the nucleon (Eq. \ref{deltas}) has been extracted from $g_{1}$ analysis under the assumption of SU(3) symmetry and found to be negative ($2\Delta S=-0.09 \pm 0.01 \pm 0.02$ from \cite{Alekseev:2007vi}) by several high energy scattering experiments.

\vspace{-5pt}
\begin{equation}
\int_0^1 [\Delta s(x) + \Delta\bar{s} (x)]dx=2\Delta S
\label{deltas}
\end{equation}

\begin{wrapfigure}{r}{0.35\textwidth}
\vspace{-30pt}
\begin{center}
\includegraphics[width=0.33\textwidth]{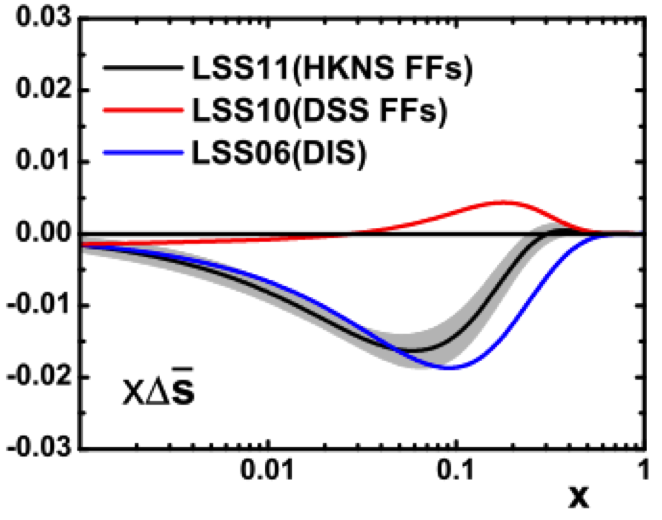}
\end{center}
\vspace{-29pt}
\caption{Polarisation of strange quarks from \cite{Leader:2011tm}.}
\label{DeltaS}
\end{wrapfigure}
 
An alternative approach to measure $\Delta S$ is in semi-inclusive DIS by measuring longitudinal spin asymmetries for identified final-state charged hadrons and assuming the knowledge of FFs which play a key role. In this case, a small positive value of $\Delta S$ was found by the HERMES experiment \cite{Airapetian:2008qf} with $2\Delta S= 0.037 \pm 0.019 \pm 0.027$ and another null value (within errors) was found by the COMPASS experiment  \cite{Alekseev:2010ub}. However the extraction of $\Delta s(x)$ using semi-inclusive deep-inelastic scattering data strongly depends on the choice of the FFs which differ from one QCD-based fit to the other. This is illustrated in Fig.\ref{DeltaS} which shows a comparison between three different QCD-based fits of $x\Delta \bar{s}(x)$ by E. Leader \textit{et al.}~\cite{Leader:2011tm}: LSS06 fit based on purely inclusive DIS data sets, LSS10 and LSS11 based on both inclusive and semi-inclusive DIS data sets and differing by the choice of FFs. LSS11 uses the HKNS FFs based on e$^+$e$^-$ data while LSS10 uses DSS FFs based on e$^+$e$^-$, $pp$ and SIDIS data. A more accurate and precise determination of FFs is essential to solve this $\Delta S$ puzzle.

\section{Single-hadron Multiplicities: Data Analysis \& Results}

The analysis is based on a data set recorded by the COMPASS experiment in 2006 and includes event selection, particle identification (PID), determination of and correction for experimental and kinematic limitations of the apparatus as well as particle identification inefficiency. Selected deep-inelastic scattering events are required to have a reconstructed interaction vertex inside the fiducial target volume, associated to an incident and a scattered muon tracks. The incident muon energy is constrained to the range $140 < E_{\mu} <180$ GeV. Events are accepted if the photon virtuality $Q^{2} > 1$ (GeV/c)$^2$ and the energy of final-state hadronic system $W > 5$ GeV/c. These requirements select the DIS regime and exclude the nucleon resonances region. The fractional energy transferred from the incident muon to the virtual photon is constrained to the range $0.1 < y < 0.7$ to exclude the kinematic region where the momentum resolution degrades and the region the most affected by radiative effects respectively. All hadrons produced in the selected DIS sample are considered. Selected hadron tracks are required to have fractional energies in the interval $0.2 \le z \le 0.85$ to avoid any contamination from the target remnant fragmentation and to exclude muons wrongly identified as hadrons respectively. To ensure a full particle identification efficiency by the RICH detector, hadrons are required to have momenta in the range $10 \le P \le 40$ GeV. Hadron tracks are subject to geometric cuts: they must hit tracking detectors before/after the first magnet, should not cross more than 15 radiation lengths and the polar angle, defined with respect to the direction of the scattered muon at the RICH detector entrance, must be in the interval $10 \le \theta \le 120$ mrad. The correction for the RICH efficiency-misidentification probabilities is performed by extracting the yields for particle of true type from the measured yields:

\begin{equation}
N_{i}^{true} = \sum_{j} P^{-1}_{ij} \cdot N_{j}^{meas} 
\label{Relation1}
\end{equation}

Here $P^{-1}$ is the inverse of the RICH PID probability matrices with $i=$($\pi, K, p$). Diagonal elements represent efficiencies and off-diagonal elements represent misidentification probabilities. The elements of the $3 \times 3$ $P$ matrix,  which depend on momentum and polar angle of final-state hadron track, are evaluated from experimental data. For pions and kaons, the PID efficiencies are extracted using the decay of $\phi \rightarrow K^+ K^-$ and of $K_{S} \rightarrow \pi^+ \pi^-$ respectively. For protons, the decay of $\Lambda \rightarrow p \pi^-$ and its charge-conjugate are used. The PID matrix elements are constrained by $\sum_{h'} P^{h \rightarrow h'} < 1$.

Experimental multiplicities are corrected for the limited geometric and kinematic acceptance of the apparatus as well as for detector inefficiencies using a Monte-Carlo (MC) simulation of $\mu - N$ scattering reaction. Events are generated using LEPTO~\cite{Ingelman:1996mq} lepton-nucleon DIS generator where the hadronisation mechanism is simulated with JETSET package~\cite{Sjostrand:1995iq}. Although the acceptance is generator-independent, a dedicated tuning of the hadronisation parameters was used to ensure reasonable MC description of data. The response of the spectrometer is simulated with a GEANT3-based program and MC data are reconstructed with the same software as the experimental data. The acceptance correction factor is differentially estimated, for each hadron charge and type, as a function of $x$, $y$ and $z$. In each kinematic bin ($x$,$y$,$z$) it is defined as:

\begin{equation}
 A_{h}(x^{\rm{rec}},y^{\rm{rec}},z^{\rm{rec}}) = \frac{M_{h}^{\rm{rec}}(x^{\rm{rec}},y^{\rm{rec}},z^{\rm{rec}}) }{M_{h}^{\rm{gen}} (x^{\rm{gen}},y^{\rm{gen}},z^{\rm{gen}})}
\end{equation}

where $M_{h}^{\rm{rec}}$ ($M_{h}^{\rm{gen}}$) is the reconstructed (generated) hadron multiplicity in the reconstructed (generated) values of kinematic variables $x$, $y$ and $z$. While the MC reconstructed hadron multiplicity is subject to kinematic and geometrical cuts, only kinematic cuts on $Q^2$, $x$, $y$, $z$ and $P$ are applied to the MC generated hadron multiplicity. Experimental hadron multiplicities are corrected for radiative effects as described in \cite{Makke:2011fia}. The COMPASS isoscalar target consists of granulated $^6$LiD immersed in liquid helium. The small admixtures of H, $^3$He and $^7$Li lead to an excess of neutrons of about 0.1\%. A study of possible nuclear effects has been performed by comparing data taken on $^6$LiD with data recorded on liquid hydrogen target (NH$_3$) and no effect has been observed. Finally, hadrons originating from the decay of diffractively produced vector mesons are not excluded from the sample.

\section{Results}

\begin{figure}[htdp]
\centering
\includegraphics[height=10.cm,width=.9\textwidth]{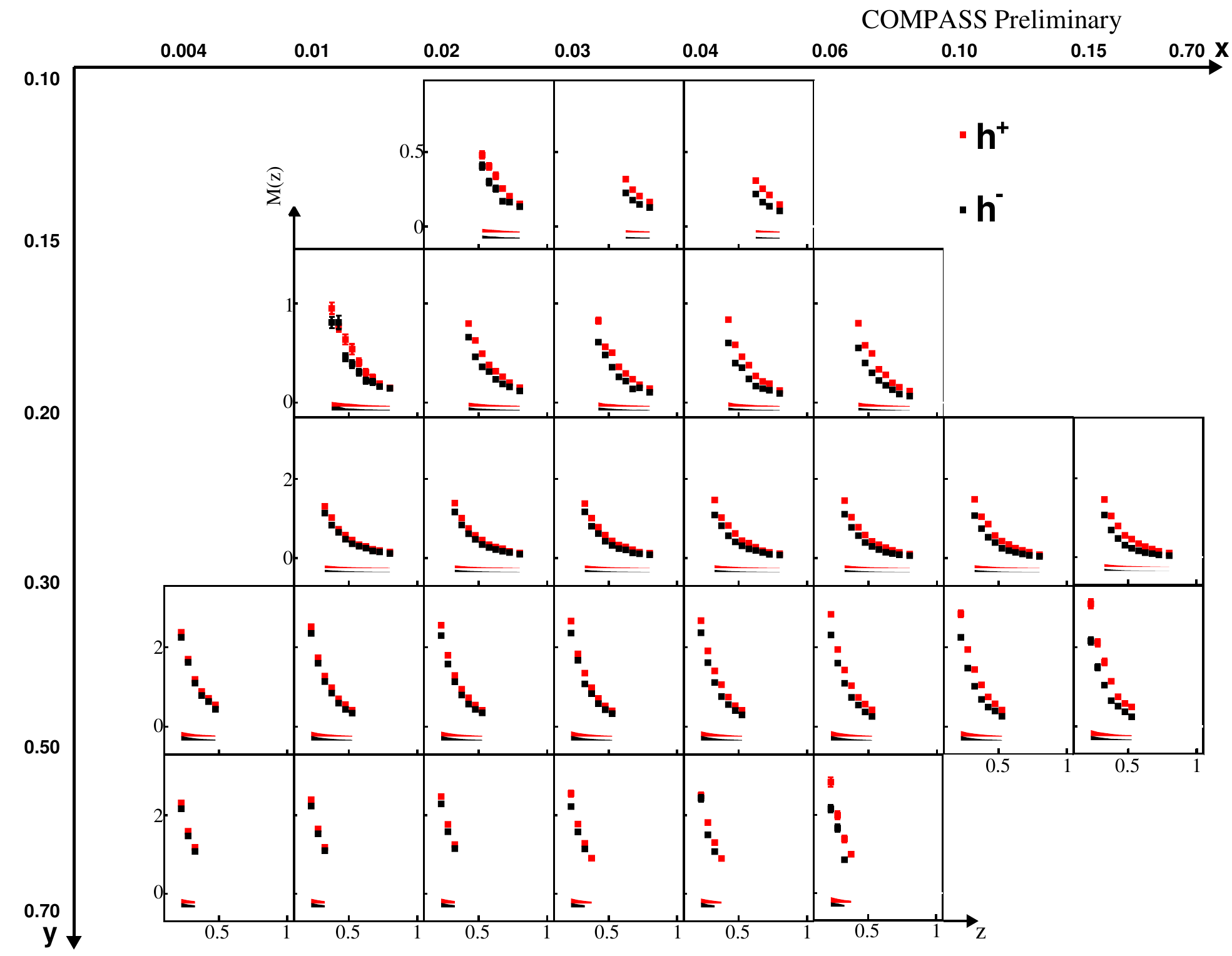}
\caption{Charged hadron multiplicities.}
\label{Mulh}
\end{figure}

\begin{figure}[htdp]
\centering
\subfigure[Pions  ]{\label{}\includegraphics[height=10.cm,width=.9\textwidth]{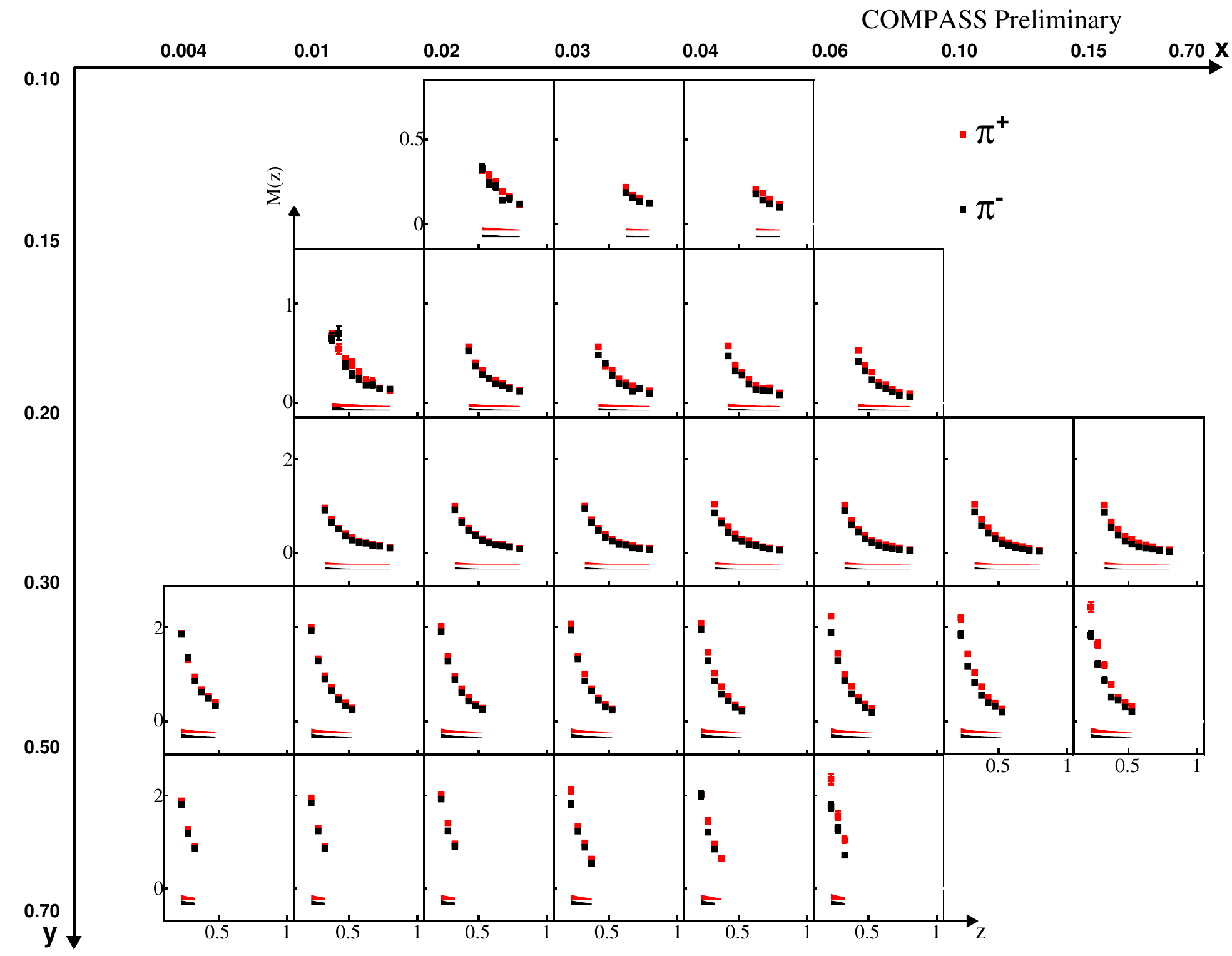}}
\subfigure[Kaons]{\label{}\includegraphics[height=10.cm,width=.9\textwidth]{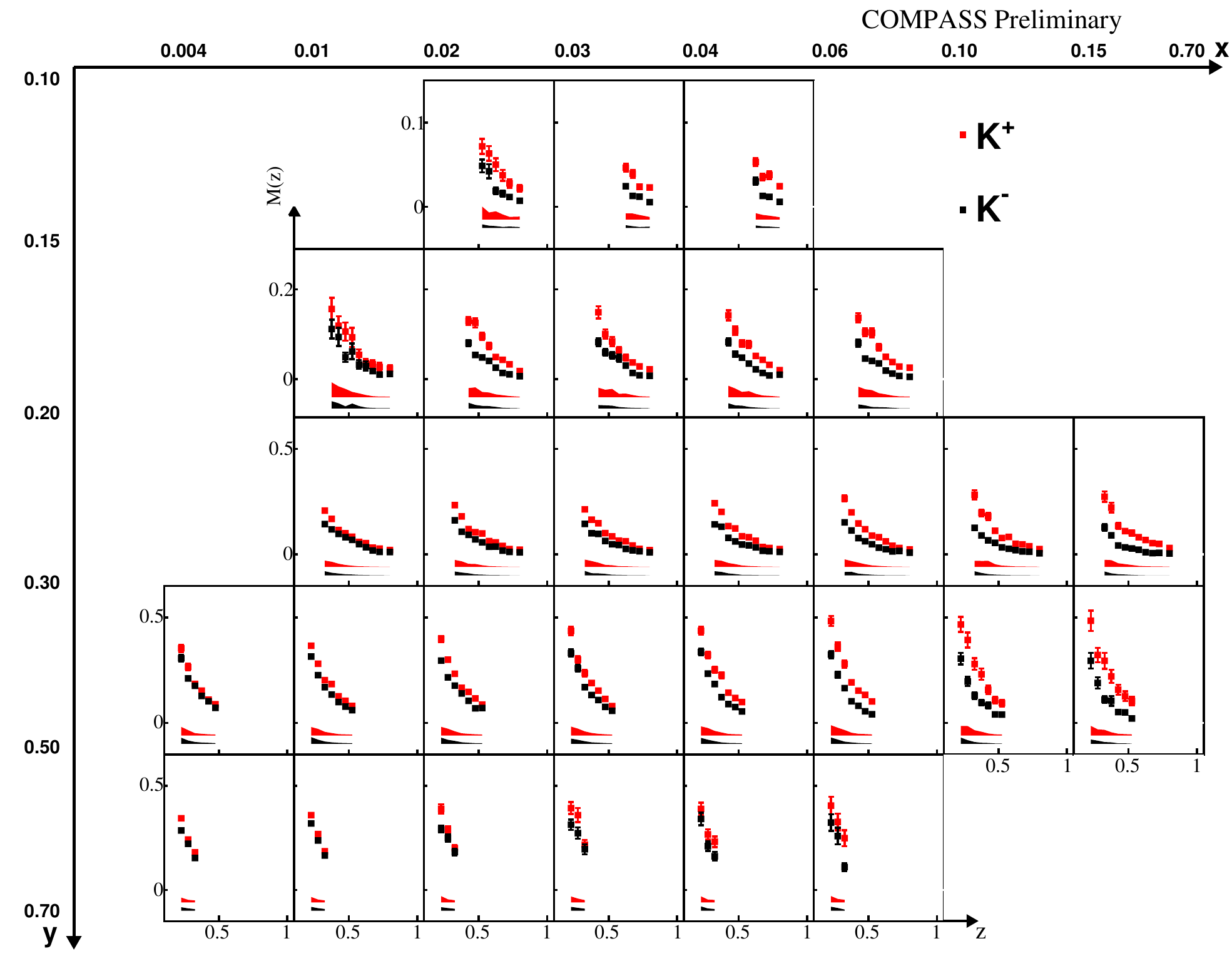}}
\caption{Charged pion (top) and kaon (bottom) multiplicities.}
\label{Mulxyz}
\end{figure}

Figs.~\ref{Mulh} and \ref{Mulxyz} show differential multiplicities for charged hadrons, pions and kaons $\textit{vs.}$ $x$, $y$ and $z$. The bands show the systematic uncertainties to which different sources of systematics contribute: PID RICH performance, acceptance evaluation, time and MC model dependence. Note that systematic uncertainties are fully correlated at low $y$ and low $z$. 

Another interesting quantity to study is the integrated sum of charged kaon multiplicities obtained from the ($x$,$y$,$z$) dependent multiplicities by averaging over $y$ and integrating over $z$ in the measured range [$0.2, 0.85$]. It reflects the contribution of the nucleon strange quarks to kaon production via

\begin{equation}
\int M^{K^+ + K^-}(x,z) dz = \frac{Q(x)\int D_{Q}^{K}(z)dz + S(x)\int D_{S}^{K}(z)dz}{5Q(x) + 2S(x)}
\label{sum}
\end{equation}

\begin{figure}[htdp]
\centering
\includegraphics[height=6.cm,width=.6\textwidth]{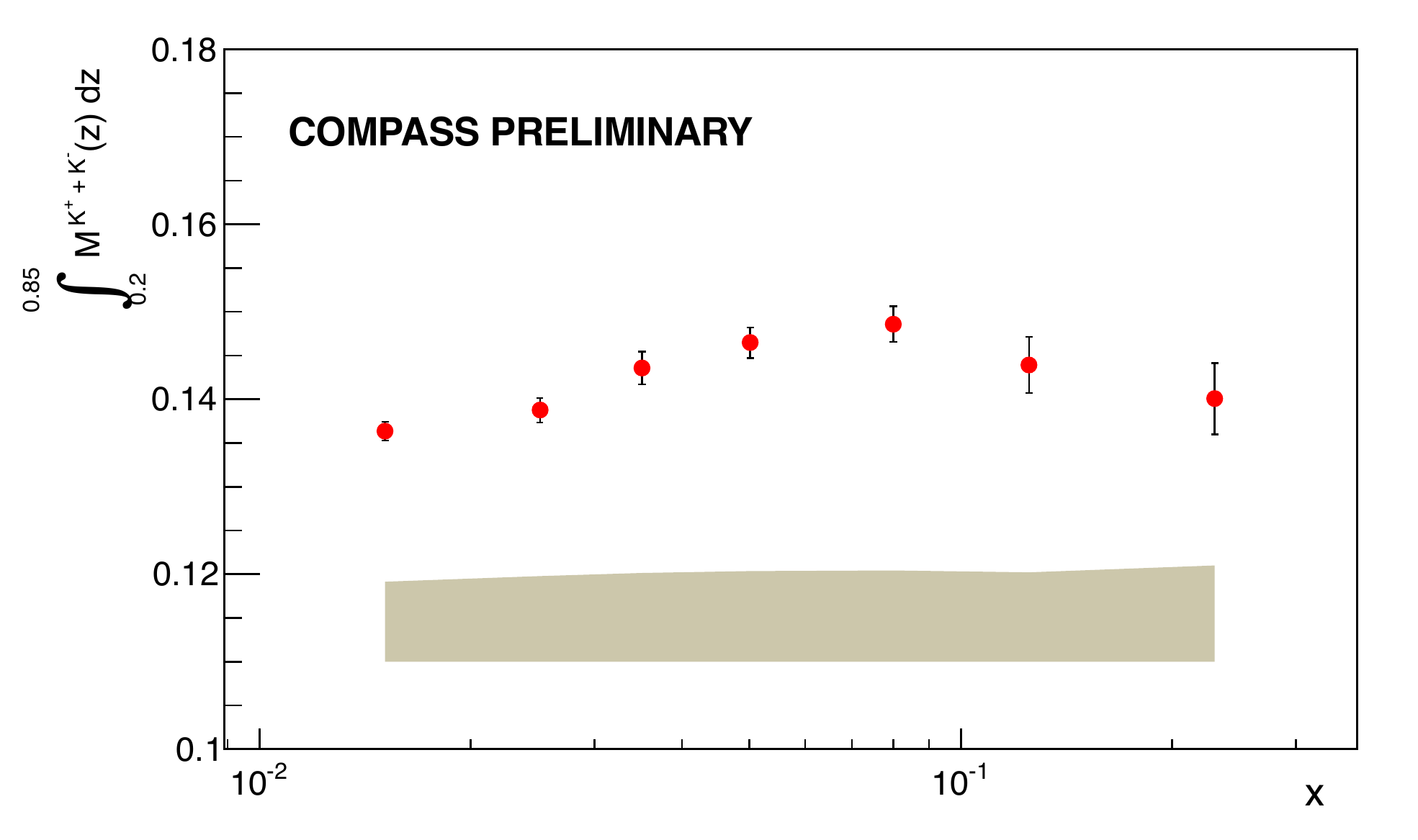}
\caption{$z$-Integrated sum of K$^{\pm}$ multiplicities.}
\label{Sum}
\end{figure}

Here $Q(x) = u(x) + \bar{u}(x) + d(x) + \bar{d}(x)$, $S(x)=s(x) + \bar{s}(x)$, $D_{Q}^{K} = 4D_{fav}^{K} + 6D_{unf}^{K}$ and $D_{S}^{K} \simeq 2D_{str}^{K}$ where $D_{fav}$, $D_{unf}$ and $D_{str}$ denote favoured ($D_{u}^{K^+}$), equally unfavoured ($D_{d}^{K^+} = D_{\bar{d}}^{K^+} = D_{s}^{K^+}=D_{\bar{u}}^{K^+}$) and strange ($D_{\bar{s}}^{K^+}$) kaon FFs respectively. 
The importance of the integrated charged kaon sum is in its behavior at the highest and the lowest $x$ regions. At high $x$,  $S(x)\simeq 0$ and the sum consequently equals a constant ($\simeq \frac{1}{5}\int D_{Q}^{K}(z)dz$).  At low $x$, the non-strange term slightly decreases and the strange term is non-null. The sum thus becomes sensitive to the strangeness component of the nucleon, i.e. ($\simeq \frac{1}{5}[\int D_{Q}^{K}(z)dz + \frac{S(x)}{Q(x)}\int D_{S}^{K}(z)dz]$).  In this region, significant increase of the sum would suggest important contribution of the strange quarks via the product $S(x)\int D_{S}^{K}(z)dz$. Fig.~\ref{Sum} shows the integrated sum of charged kaons extracted from COMPASS data.  At high $x$, COMPASS data follow a constant within systematic uncertainties as expected while at low $x$, no strong growth of the sum is observed. This rather suggest a small value of the strange product $S(x)\int D_{S}^{K}(z)dz$.  Further study of kaon multiplicities is ongoing in order to improve the systematic uncertainty shown by the band in Fig. \ref{Sum}.

\vspace{-0.2cm}
\section{Summary}
COMPASS has performed a precision measurement of the multiplicity of charged pions and kaons produced in semi-inclusive DIS on an isoscalar target as a function of $x$, $y$ and $z$ in the perspective of contributing to the QCD-based fit of FFs. The measurement covers a wide kinematic range  defined by $Q^{2} > 1$ (GeV/c)$^2$, $0.1 < y < 0.7$, $4. 10^{-3} < x < 0.7$ and $0.2 < z < 0.85$ and provides a valuable input for global QCD-based determinations of FFs. An analog analysis based on the 2012 data set recorded using a proton target is starting.


\begin{thebibliography}{99}
\vspace{-0.3cm}
\itemsep -4pt

\bibitem{Adolph:2012nw}
  C.~Adolph {\it et al.}  [COMPASS Collaboration],
  \href{http://www.sciencedirect.com/science/article/pii/S0370269312005321}{Phys.\ Lett.\ B {\bf 713} (2012) 10}
  \href{http://arxiv.org/abs/arXiv:1202.6150}{arXiv:hep-ex/12026150}.

\bibitem{MakkeSPIN12}
  N.~Makke
  contribution given at the 20th International Symposium on Spin Physics (SPIN2012) 
  
\bibitem{Alekseev:2007vi}
  M.~Alekseev {\it et al.}  [COMPASS Collaboration],
    \href{http://www.sciencedirect.com/science/article/pii/S0370269308000427}{Phys.\ Lett.\ B {\bf 660} (2008) 458}\href{http://arxiv.org/abs/arXiv:0707.4077}{arXiv:hep-ex/07074077}.
  
\bibitem{Airapetian:2008qf}
  A.~Airapetian {\it et al.}  [HERMES Collaboration],
  \href{http://www.sciencedirect.com/science/article/pii/S0370269308009386}{Phys.\ Lett.\ B {\bf 666} (2008) 446}, \href{http://arxiv.org/abs/0803.2993}{arXiv:hep-ph/08032993}

\bibitem{Alekseev:2010ub}
  M.~G.~Alekseev {\it et al.}  [COMPASS Collaboration],
 \href{http://www.sciencedirect.com/science/article/pii/S0370269310009810}{Phys.\ Lett.\ B {\bf 693} (2010) 227}, \href{http://arxiv.org/abs/arXiv:1007.4061}{arXiv:hep-ph/10074061}.

\bibitem{Leader:2011tm}
  E.~Leader, A.~V.~Sidorov and D.~B.~Stamenov,
 \href{http://prd.aps.org/abstract/PRD/v84/i1/e014002}{Phys.\ Rev.\ D {\bf 84} (2011) 014002}, \href{http://arxiv.org/pdf/1103.5979.pdf}{arXiv:hep-ph/11035979}
  
\bibitem{Makke:2011fia}
  N.~Makke,
  PhD thesis, CEA/Saclay IRFU/SPhN, \href{http://inspirehep.net/record/1231209/files/2011_phd_makke.pdf}{CERN-THESIS-2011-279}.

\bibitem{Ingelman:1996mq}
  G.~Ingelman, A.~Edin and J.~Rathsman,
  \href{http://www.sciencedirect.com/science/article/pii/S0010465596001579}{Comput.\ Phys.\ Commun.\  {\bf 101} (1997) 108}, \href{http://arxiv.org/abs/hep-ph/9605286}{arXiv:hep-ph/9605286}
  
\bibitem{Sjostrand:1995iq}
  T.~Sjostrand,
  \href{http://arxiv.org/abs/hep-ph/9508391}{arXiv:hep-ph/9508391}

\end{thebibliography}
\end{document}